\documentclass[%
 reprint,
 amsmath,amssymb,
 aps,
 pra
]{revtex4-2}

\usepackage{babel}

\usepackage{graphicx}
\usepackage{dcolumn}
\usepackage{bm}
\usepackage{inputenc}
\usepackage{lipsum}
\usepackage{xcolor}
\usepackage{tikz}
\usepackage{enumerate}
\usetikzlibrary{arrows.meta,calc} 
\makeatletter
\def\l@en{\l@english}
\makeatother
\usepackage{physics}

\begin{document}

\preprint{APS/123-QED}

\title{Characterization of the multimode nature of single-photon sources based on spontaneous parametric down conversion}

\author{Emil R. Hellebek}
 
\author{Klaus Mølmer}
\author{Anders S. Sørensen}
\affiliation{Center for Hybrid Quantum Networks (Hy-Q), Niels Bohr Institute, University of Copenhagen, Blegdamsvej 17, Copenhagen DK-2100, Denmark}

\date{\today}

\begin{abstract}

Single-photon sources are necessary components for many prospective quantum technologies. One candidate for a single-photon source is spontaneous parametric down conversion combined with a heralding photon detection. 
The heralded light pulse from such a source, is typically treated as single-mode, this treatment, however, is incomplete. We develop a full multimode description based on the exact Bogoliubov treatment of the down conversion process. We then provide a perturbative and effective treatment, which illustrates the most important physical mechanisms and permits analytical estimates of the success probability and purity of single-photon states under practical heralding conditions, both without relying on the precise detection time of the heralding photon and when accepting photons only in a narrow window around the time of the detection. This permits us to characterize the emitted light under three different assumptions for the pump pulse. For spontaneous parametric down conversion with a very short pump pulse, we find the single-mode description to be accurate, while for longer pump pulses and continuous pumping, a multimode description is necessary. Our findings can be used to guide the design of quantum information protocols based on heralded single-photon sources, as their performance may depend on the multimode nature of the sources.

\end{abstract}

\maketitle
\section{Introduction}

Single-photon sources are crucial components in the advancement of many of the prospective quantum technologies under the label of the second quantum revolution \cite{benyoucef_second_2023,kimble_quantum_2008}. This includes secure communication through device-independent quantum cryptography \cite{gisin_quantum_2002,yuan_entangled_2010}, quantum computation \cite{browne_resource-efficient_2005} and the establishment of long-distance quantum networks \cite{sangouard_long-distance_2007}. The performance of these technologies depends heavily on the performance of the single-photon source. Characterizing the single photon source is thus essential for quantifying the performance of any setup incorporating the source. This characterization, furthermore, helps illuminate some of the possible choices when designing protocols using single-photon sources.

A candidate for single-photon sources employs an Optical Parametric Oscillator (OPO). In an OPO, the process of spontaneous parametric down conversion (SPDC) converts pump photons of frequency $\omega$ into photon pairs at frequencies $\omega_A$ (the signal photon) and $\omega_B=\omega-\omega_A$ (the idler photon). Subject to a classical pump, this process generates a superposition of quantum states with zero, one, and more photon pairs; hence, the detection of an idler photon heralds the presence of one or more signal photons. When the OPO is driven weakly, an approximate single-photon state is heralded by the detection event \cite{hong_theory_1985,hong_experimental_1986,kaneda_heralded_2016}. Various methods of driving an OPO have been employed in experiments, including short pulses conditioned on a detection at any time \cite{mosley_heralded_2008,zhang_preparation_2011}, and continuous driving, where the heralded signal photon occupies a wave packet mode centered around the time of the heralding event \cite{neergaard-nielsen_generation_2006,rielander_cavity_2016}.

The SPDC process is generally treated as single-mode \cite{gerry_introductory_2005}. To fully characterize the heralded quantum state of the light, a multimode description is, however, in general required \cite{neergaard-nielsen_generation_2006,tualle-brouri_multimode_2009,nielsen_multimode_2007,nielsen_photon_2007,molmer_non-gaussian_2006,kopylov_theory_2024,christ_probing_2011,quesada_beyond_2022,sonoyama_analysis_2022}. In this study, we present such a multimode description of the light from an OPO based on the exact Bogoliubov treatment of the down conversion process. From this, we can provide a perturbative treatment, which allows us to characterize the light under different experimental implementations, allowing us to estimate important quantities for the performance of the OPO as a heralded single-photon source under different circumstances.

The heralded single-photon source can be employed with great benefit in the creation of the long distance network \cite{simon_quantum_2007}, by mixing the idler beams of two of these sources on a beamsplitter before the detection. A detection of a photon will thus herald the creation of a superposition of a photon being present in either of the signal arms. In this paper, we consider limits appropriate for employing the source in a long distance quantum network, however, the results can easily be generalized to other applications.

The paper is structured as follows: 
In Section \ref{sec:model}, we apply input-output theory to describe the fields produced by an OPO and determine the temporal correlations within the fields. We furthermore describe the effect on the signal field by the detection of an idler photon.
In Section \ref{sec:charchter} we define the characterization parameters used, to quantify the quantum nature of the signal light field.
In Section \ref{sec:Results}, we apply our theory to different experimental implementations of pulsed and continuously driven OPO-based single-photon sources. 
Section \ref{sec:Conclusion} summarizes the conclusions of our study.

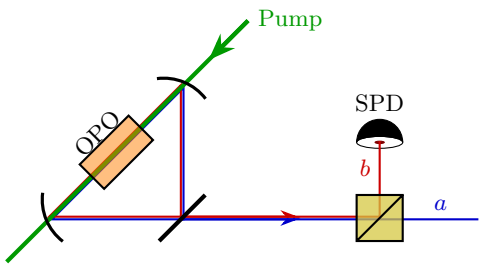
\begin{figure}
    \centering
    \begin{tikzpicture}[scale=1.75,
        pump/.style={ultra thick,green!60!black},
        signal/.style={thick,blue!80!black,transform canvas={yshift=-0.5}},
        idler/.style={thick,red!80!black,transform canvas={yshift=0.5}},
        BS/.style = {thick,fill = cyan,fill opacity =0.75}
        ]
        \pgfmathsetmacro{\cav}{1}
        \pgfmathsetmacro{\dbs}{3*\cav/2}
        \pgfmathsetmacro{\ddet}{3/4*\cav}

        \pgfmathsetmacro{\BS}{0.175}
        \pgfmathsetmacro{\detrad}{\BS}
        \pgfmathsetmacro{\Detang}{15}
        
        \pgfmathsetmacro{\ms}{0.4}
        \pgfmathsetmacro{\OPOx}{3/2*\BS}
        \pgfmathsetmacro{\OPOy}{\OPOx/2}
        
        \coordinate (M1) at (0,0);
        \coordinate (M2) at (0,\cav);
        \coordinate (M3) at (-\cav,0);
        \coordinate (Start) at ($(M2)+(\cav/2,\cav/2)$);
        \coordinate (OPO) at (-\cav/2,\cav/2);
        \coordinate (BS) at (\dbs,0);
        \coordinate (det) at (\dbs,+\ddet-\detrad);

        \draw[fill=black] ($(det)+(\detrad,0)$) arc (0:180:\detrad) node[above,pos=0.5] {SPD} arc  (180:0:0.175 and sin{\Detang}*0.175);
        \draw (det) circle [x radius = \detrad, y radius = \detrad*sin{\Detang}];
        \filldraw[color=red!50!black] (det) circle [x radius = \detrad/5, y radius = \detrad*sin{\Detang}/5] ;

        \draw[signal,transform canvas={xshift=0.5}] (M3) -- (M1) -- (M2) -- cycle;
        \draw[idler,transform canvas={xshift=-0.5}] (M3) -- (M1) -- (M2) -- cycle;
        \draw[signal] (M1) -- (BS) -- ++(\ddet,0) node[above,pos=1/2*(1+\BS/\ddet)] {$a$};
        \draw[signal,arrows={-Stealth[harpoon,swap]}] (M1) -- ($(M1)!0.6!(BS)$);
        \draw[idler,arrows={-Stealth[harpoon]}] (M1) -- ($(M1)!0.6!(BS)$);
        \draw[idler] (M1) -- (BS) -- (det) node [left,pos=1/2*(1+\BS/(\ddet-\detrad))] {$b$};
        
        \draw[pump] (Start) -- (M2) node[pos=0,right] {Pump} -- (M3)--++(-\cav/3,-\cav/3) ;
        \draw[pump,arrows={-Stealth}] (Start) -- ($(Start)!0.6!(M2)$);
        
        \draw[ultra thick] (M1) -- ++(\BS,\BS) (M1) -- ++(-\BS,-\BS);
        \draw[BS,fill=yellow!80!black] (BS) -- ++(-\BS,-\BS)  rectangle ++(2*\BS,2*\BS) -- ++(-\BS,-\BS);

        \draw[very thick] ([xshift=sin{22.5},yshift=cos{22.5}]M2) arc (90-22.5:120-22.5:\ms)([xshift=sin{22.5},yshift=cos{22.5}]M2) arc (90-22.5:60-22.5:\ms);
        \draw[very thick] ([xshift=-cos{22.5},yshift=-sin{22.5}]M3) arc (90+112.5:120+112.5:\ms)([xshift=-cos{22.5},yshift=-sin{22.5}]M3) arc (90+112.5:60+112.5:\ms);
        
        \draw[thick,rotate = 45,fill=orange,fill opacity=0.5] ($(OPO)-(\OPOx,\OPOy)$) rectangle ($(OPO)+(\OPOx,\OPOy)$);
        \node (label) at ($(OPO)+(-1.1*\OPOy,1.1*\OPOy)$) [rotate =45] {OPO};
    \end{tikzpicture}
    \caption{
    Schematic illustration of the setup. A classical pump laser beam (green line) drives an OPO (orange box), generating photon pairs in the non-degenerate signal (blue line) and idler (red line) cavity modes. The light exits the cavity at a rate $\kappa$, producing continuous signal ($a(t)$) and idler ($b(t)$) fields, which are separated by a dichroic mirror (yellow box). The idler beam impinges on a single-photon detector (SPD).
    } 
    \label{fig:Sketch}   
\end{figure}

\section{Theoretical model of the heralded single-photon source\label{sec:model}}
The system under investigation is schematically illustrated in Fig. \ref{fig:Sketch}. Within a cavity featuring a full width half max linewidth $\kappa$, a pump beam interacts with a non-linear crystal, generating pairs of photons in the intra-cavity modes $a_c$ and $b_c$. The two cavity fields exit the cavity and combine with incident vacuum fields $a_\text{in}$ and $b_\text{in}$ to form the output multimode signal and idler fields $a$ and $b$. These fields are split by a dichroic mirror, and the detection of a photon in the idler beam is accomplished using a single-photon detector (SPD), which heralds the presence of a photon in the signal beam.

\subsection{Input-output relations and mode decomposition \label{sec:mode}}
We begin by detailing how the input fields are transformed by the OPO into the corresponding output fields. Our results in this section depends on the specific scenario, but they can be generalized with small modifications. The Hamiltonian describing the SPDC process in a cavity is \cite{hong_theory_1985}
\begin{align}
    \mathcal H = \hbar\chi(t) (a_cb_c+a_c^\dagger b_c^\dagger),
\end{align} 
where $\chi(t)$ is given by the temporal profile of the pump pulse.

The dynamics of the cavity fields are given by the Heisenberg equations of motion, and the output fields are governed by the input-output relations \cite{gardiner_input_1985}, resulting in 
\begin{subequations}\label{eq:EOM+in}
    \begin{align}\label{eq:EOM}
        \dot {\vb{w}}_c &= -\frac{\kappa}{2} \vb{w}_c +\sqrt{\kappa}\, \vb{w}_\text{in} + \tau_y \chi(t) \vb{w}_c,\\
        \vb{w} &= \vb{w}_{\text{in}}-\sqrt{\kappa}\,\vb{w}_c
    \end{align}
\end{subequations}
where $\vb{w}_i=(a_i,\, b_i^\dagger)^T$ and $\tau_y$ is the Pauli-$y$ matrix introduced for brevity. 

As the commutation relations holds for all the fields, the solution of the differential equation can be written of the form of a general Bogoliubov transformation, with exchange symmetry between the $a$ and $b$ fields
\begin{align}\label{eq:bog}
    \vb{w}(t) = \int_{-\infty}^\infty dt' \begin{pmatrix}
        u(t,t') & v(t,t')\\
        v^*(t,t') & u^*(t,t')
    \end{pmatrix}\vb{w}_\text{in}(t'),
\end{align}
where $u(t,t')$ and $v(t,t')$ are the Bogoloiubov transformation functions. From the commutation relations, we obtain identities obeyed by $u(t,t')$ and $v(t,t')$
\begin{subequations}\label{eq:bogrels}
    \begin{align}\label{eq:bogrela}
        \delta(t-t')&=\int_{-\infty}^{\infty} d\tau  \left[u(t,\tau)u^*(t',\tau)-v(t,\tau)v^*(t',\tau) \right]\\\label{eq:bogrelc}
        0 &= \int_{-\infty}^{\infty} d\tau  \left[u(t,\tau)v(t',\tau)-v(t,\tau)u(t',\tau) \right],
    \end{align}
\end{subequations} 
By solving the first order differential equation \eqref{eq:EOM+in}, we identify the expressions for $u(t,t')$ and $v(t,t')$
\begin{subequations}\label{eq:bogtrans}
    \begin{align}
        u(t,t') &= \delta(t-t') -\kappa \Theta(t-t') e^{-\frac{\kappa}{2}(t-t')}\cosh[\mathcal{I}(t,t')]\\
        v(t,t') &= i\kappa \Theta(t-t') e^{-\frac{\kappa}{2}(t-t')}\sinh[\mathcal{I}(t,t')],
    \end{align}
\end{subequations}
where $\mathcal{I}(t,t')=\int_{t'}^t d\tau \chi(\tau)$ and $\Theta(t)$ is the Heaviside step function.

We will now follow the method outlined in Refs. \cite{christ_probing_2011,quesada_beyond_2022} to find the state of the light exiting the cavity. We make a singular value decomposition of the transformation functions $u(t,t')$ and $v(t,t')$. The results here rely on the exchange symmetry between the output fields $a$ and $b$. For a more general derivation  see Appendix \ref{sec:AppBog}.
According to the Bloch-Messiah Theorem for bosons \cite{braunstein_squeezing_2005}, the left and right singular vectors of the two transformation functions are related such that
\begin{subequations}
    \begin{align}
        u(t,t') &= \sum_\ell \lambda_\ell f_\ell(t) g^*_\ell(t')\\
        v(t,t') &= \sum_\ell \mu_\ell f_\ell(t) g_\ell(t'),
    \end{align}
\end{subequations}
and the singular values are related through 
\begin{align}
    \lambda_\ell^2-\mu_\ell^2=1,
\end{align}
which allows us to write $\lambda_\ell = \cosh(\xi_\ell)$ and $\mu_\ell=\sinh(\xi_\ell)$.

Rewriting Eq. \eqref{eq:bog} using the singular value decomposition we find
\begin{align}
    \vb{w}(t) = \sum_\ell \begin{pmatrix}
        \cosh(\xi_\ell) f_\ell(t) & \sinh(\xi_\ell) f_\ell(t)\\
        \sinh(\xi_\ell) f_\ell^*(t) & \cosh(\xi_\ell) f_\ell^*(t) 
    \end{pmatrix}
    \vb{w}_{\text{in},\ell},
\end{align}
where $\vb{w}_{\text{in},\ell}^T=(a_{\text{in},\ell},b_{\text{in},\ell}^\dagger)$ is the vector containing the input fields in the mode shape described by $g_\ell(t)$, i.e. $a_{\text{in},\ell}=\int dt' g_\ell^*(t')a_\text{in}(t')$ and $b_{\text{in},\ell}^\dagger=\int dt' g_\ell(t')b_\text{in}^\dagger(t')$. We can similarly transform the output fields to the basis described by $f_\ell(t)$ through $a_{\ell} = \int dt f_\ell^*(t) a(t)$ and $b_{\ell}^\dagger = \int dt f_\ell(t) b^\dagger(t)$. This yields the transformation
\begin{align}
    \vb{w}_\ell = \begin{pmatrix}
        \cosh(\xi_\ell) & \sinh(\xi_\ell) \\
        \sinh(\xi_\ell) & \cosh(\xi_\ell) 
    \end{pmatrix} \vb{w}_{\text{in},\ell}.
\end{align}
This transformation can equivalently be written using the two-mode squeezing operator
\begin{align}
    \vb{w}_\ell = S_{\ell}^\dagger(\xi_\ell) \vb{w}_{\text{in},\ell}S_{\ell}(\xi_\ell),
\end{align}
where $S_\ell(\xi_\ell) = \exp[\xi_\ell( a_\ell^\dagger b_\ell^\dagger-a_\ell b_\ell)]$. The light in the mode shaped by $f_\ell(t)$ is thus described by the two-mode squeezed state $\ket{\text{TMSS}_\ell}$, which can be expanded in the Fock states $\ket{n_\ell,n_\ell}$ of the signal and idler mode functions
\begin{align}\label{eq:TMSS}
    \ket{\text{TMSS}_\ell} = \frac{1}{\cosh{(\xi_\ell)}} \sum_n \tanh^n(\xi_\ell) \ket{n_\ell,n_\ell}.
\end{align}
The combined state of the light exiting the cavity can thus be described as a product of two-mode squeezed states in the different modes shaped by the left singular vectors, i.e. $\ket{\Psi}=\bigotimes_\ell \ket{\text{TMSS}_\ell}$.
Thus, by choosing the correct basis, we can describe the light exiting the cavity as a product of two mode squeezed states in independent orthogonal modes.

\subsection{Correlations and the conditional density matrix after heralding \label{sec:densitymatrix}}

We now turn to the description of the quantum state of the signal beam heralded by the detection of a photon in the idler beam. The description is independent of the particular down-conversion process investigated above, and the results from this section are thus valid for any photon pair source.

We will be working in a regime where the OPO is driven weakly, such that the success probability of detecting photons in the idler beam, $P_S$, is small. We will thus expand all the relevant quantities perturbatively in $P_S$. 

The heralding click in the SPD corresponds to the annihilation of a photon in the idler beam. We shall assume that the collection efficiency of the idler beam $\eta_d$ is small, which is the case when considering a heralded single photon source employed in long distance quantum networks. Multiple detection events can thus be ignored, and we can write 
\begin{align}\label{eq:PS}
    P_S = \eta_d \int_\mathcal{T} dt \ev{b^\dagger(t) b(t)},
\end{align}
where the detector signal is recorded in the time interval $\mathcal{T}$.  We shall from now on ignore $\eta_d$ for brevity, noting that all success probabilities should be multiplied by $\eta_d$.

A detection event at time $t_c$ heralds the preparation of the signal beam reduced density matrix 
\begin{align}\label{eq:rho_trans} 
    \hat \rho \to \hat\rho_{A|t_c}=\frac{\Tr_B[b(t_c)\hat \rho b^\dagger(t_c)]}{\ev{b^\dagger(t_c) b(t_c)}},
\end{align}
where $\text{Tr}_B$ is the trace over the idler-photon subspace.

To develop our formalism, we partition the conditional reduced density matrix in the signal arm into components with a definite number of photons, i.e.
\begin{align}\label{eq:rhoaexp}
    \hat \rho_{A|t_c} = \bigoplus_n P_{n}\hat \rho_{n|t_c},
\end{align}
where $\hat \rho_{n|t_c}$ denotes the normalized density matrix with $n$ photons in the signal beam and $P_{n}$ is the population of that state.

We rewrite the density matrix components explicitly in the temporal representation
\begin{align}\label{eq:varrho_gen}
    \hat \rho_{n|t_c} = \int_{\mathcal{T}} \frac{dt_1\cdots dt_{2n} }{n!^2} \rho_{n|t_c} (\{t\})\dyad{t_1,\dots t_n}{t_{n+1},\dots,t_{2n}},
\end{align}
where $\rho_{n|t_c} (\{t\})$ is the temporal representation of the density matrix, $\{t\}=\{t_1,\dots,t_{2n}\}$ and $\ket{t_1, \dots ,t_n} =a^\dagger(t_1)\cdots a^\dagger(t_n)\ket{\emptyset}$. 

We now turn to the $2n$-point correlation functions $C_{2n|t_c}(\{t\})$ of the signal beam after the heralding event, given by
\begin{align}\label{eq:c2n}
    C_{2n|t_c}(\{t\}) = \Tr[a(t_1)\cdots a(t_n)\rho_{A|t_c} a^\dagger(t_{n+1})\cdots a^\dagger (t_{2n})].
\end{align}
We can relate the density matrix components to the $2n$-point correlation functions, through the relation (see Appendix \ref{sec:App2} for derivation),
\begin{widetext}
\begin{align}\label{eq:dmgen}
    P_{n}\mel{t_1,\dots,t_n}{\hat \rho_{n|t_c}}{t_{n+1},\dots,t_{2n}} = \sum_{k=0}^\infty\frac{(-1)^k}{k!} \int_{\mathcal{T}^k} d\tau_1\cdots d\tau_k C_{2(n+k)|t_c} (\{t\}'),
\end{align}
where the extended set of time arguments is $\{t\}'=\{\tau_1,\dots, \tau_k, t_1, \dots,t_{2n},\tau_k,\dots, \tau_1$\}. 
This equation allows us to construct the full density matrix from all the even correlation functions. As the correlation functions are easy to calculate or measure compared to the density matrix, we can use this equation as a shortcut to gain information about the state.

We can write the $2n$-point signal correlation function conditioned on a specific idler click time $t_c$, by substituting the reduced density matrix, with the form from Eq. \eqref{eq:rho_trans}, yielding the expression
    \begin{align}\label{eq:cond_2n}
        C_{2n|t_c}(\{t\}) = \frac{\Tr[a(t_1)\cdots a(t_n)b(t_c)\hat \rho b^\dagger(t_c) a^\dagger(t_{n+1})\cdots a^\dagger (t_{2n})]}{\ev{b^\dagger(t_c) b(t_c)}}.
    \end{align}
\end{widetext}
Since the full emitted light from the OPO is in a Gaussian state \emph{before} the heralding event, we can apply Wicks theorem to express all higher order correlation function in terms of two-time correlations.

As we are interested in the regime, where an approximate single-photon state is heralded, we will expand $C_{2n|t_c}$ perturbatively in powers of the interaction strength. To obtain this, we will thus first expand the operators $d= \sum_j d_j$
where $d$ is either $a_c,$ $b_c,$ $a(t),$ $b(t),$ $\vb{w}_c$ or $\vb{w}$ and $d_j$ contains $\chi(t)$ $j$ times.

The set of differential equations in Eq. \eqref{eq:EOM+in} is solved recursively for each term in the series expansion, 
\begin{subequations}\label{eq:Fields}
    \begin{align}
        \vb{w}_0(t) &= \vb{w}_\text{in}(t)  - \kappa \int_{-\infty}^t dt_1e^{-\frac{\kappa}{2}(t-t_{1})}\vb{w}_\text{in}(t_{1})\\
        \vb{w}_1(t) &= -\kappa\int_{-\infty}^t dt_1\chi(t_1)\int_{-\infty}^{t_1} dt_{2} e^{-\frac{\kappa}{2}(t-t_2)}\tau_y \vb{w}_\text{in}(t_2)\label{eq:v1}\\
        \vb{w}_j(t) &= \int_{-\infty}^t dt_1\chi(t_1)e^{-\frac{\kappa}{2}(t-t_1)} \tau_y \vb{w}_{j-1}(t_1),\label{eq:vj}
    \end{align}
\end{subequations}
where Eq.~\eqref{eq:vj} holds for $j\ge 2$. 

Using the series expansion of the field operators, we obtain expressions for the two time field correlation functions in orders of the interaction strength, and define the $j$'th term in the expansion of $\ev{a^\dagger(t) a(t')}$ as
\begin{align}
    \ev{a^\dagger(t) a(t')}_j = \sum_{k=0}^j \ev{a_k^\dagger(t) a_{j-k}(t')}
\end{align}
with a similar definition for $\ev{a(t)b(t')}$.
Since the input light is in the vacuum state, we can explicitly write the leading order contributions to the output field correlation functions,

\begin{subequations}\label{eq:2points}
    \begin{align}
        \ev{a^\dagger(t) a(t')} &= \ev{a^\dagger(t) a(t')}_2 + \ev{a^\dagger(t) a(t')}_4+ \order{\chi^6}\\
        \ev{a(t)b(t')} &= \ev{a(t)b(t')}_1 + \ev{a(t)b(t')}_3 + \order{\chi^5}.
    \end{align}
\end{subequations}
The exchange symmetry between the output fields $a$ and $b$ implies $\ev*{b^\dagger(t)b(t')}=\ev*{a^\dagger (t) a(t')}$. We note from Eq. \eqref{eq:PS}, that the success probability scales as $\order{\chi^2}$, and we will thus expand up to second order in the interaction strength.
For any two times $t_i$ and $t_j$, the product of the anomalous expectation values $\ev{a(t_i)b(t_c)}\ev*{a^\dagger(t_j) b^\dagger(t_c)}$ scales as $ \order*{\chi^2}$, whereas the product of the direct expectation values $\ev{a^\dagger(t_j)a(t_i)}\ev*{ b^\dagger(t_c)b(t_c)}$ scales as $ \order*{\chi^4}$. 

We can now return to the conditional $2n$-point correlation, $C_{2n|t_c}$, in Eq.~\eqref{eq:cond_2n}. The lowest order terms in the Wick expansion of  numerator of $C_{2n|t_c}$ must contain $\ev{a(t_i)b(t_c)}\ev*{a^\dagger(t_j) b^\dagger(t_c)}$ and $n-1$ two point correlation functions of the form $\ev*{a^\dagger(t)a(t')}$. As the denominator is of the order $\order{\chi^2}$, the leading contribution to $C_{2n|t_c}$ is of the order $\order{\chi^{2n-2}}$. Consequently, $C_{2|t_c}$ and $C_{4|t_c}$ are the only correlation functions with terms at most quadratic in $\chi(t)$. Equation \eqref{eq:dmgen} thus tells us that the only density matrix components contributing to linear order in $P_S$ are $\hat \rho_{1|t_c}$ and $\hat \rho_{2|t_c}$. 
These are given by
\begin{widetext}
    \begin{subequations}\label{eq:rhos}
        \begin{align}\label{eq:rho1}
            P_1\rho_{1|t_c}(t_1;t_2)
            =&\,  \frac{\ev*{a^\dagger(t_2) b^\dagger(t_c)} \ev*{a(t_1)b(t_c)}}{\ev*{b^\dagger(t_c) b(t_c)}}\left[1-\int_\mathcal{T} d\tau \ev*{a^\dagger(\tau) a(\tau)}  \right]
            +\ev*{a^\dagger(t_2) a(t_1)} \left[1-\int_\mathcal{T} d\tau \frac{\abs{\ev*{a(\tau)b(t_c)}}^2}{\ev*{b^\dagger(t_c) b(t_c)}} \right]\\\nonumber
            &- \int_\mathcal{T} d\tau\frac{\ev*{a^\dagger(t_2) b^\dagger(t_c)} \ev*{a(\tau)b(t_c)}\ev*{a^\dagger(\tau) a(t_1)}+ \ev*{a(t_1)b(t_c)}\ev*{a^\dagger(\tau) b^\dagger(t_c)} \ev*{a^\dagger(t_2) a(\tau)}}{\ev*{b^\dagger(t_c) b(t_c)}}
            \end{align}
            \begin{align}
            \label{eq:rho2}
            P_2\rho_{2|t_c}(t_1,t_2;t_3,t_4) =&\,4
            \frac{\ev*{a^\dagger(t_3) a(t_2)}\ev*{a^\dagger(t_4) b^\dagger(t_c)}\ev*{a(t_1)b(t_c)}}{\ev*{b^\dagger(t_c) b(t_c)}}.
        \end{align}
    \end{subequations}
Using the expansion of the two point correlation functions in powers of $\chi(t)$ given in Eq. \eqref{eq:2points}, we can expand the density matrix components in powers of $\chi(t)$. To obtain the density matrix contributions to second order in $\chi(t)$, we need $\ev*{a(t)b(t')}$ up to $\order{\chi^3}$, $\ev*{a^\dagger(t) a(t')}$ up to $\order{\chi^2}$, and $\ev*{b^\dagger(t_c) b(t_c)}$ up to order $\order{\chi^4}$.
\newpage
\end{widetext}

We will consider two different possible applications of the heralding mechanism: one where we condition on the occurrence of a click, but not on the precise time of the click, and one, where we keep track of the specific detection time and the associated timing of the heralded signal photon in the state $\rho_{A|t_c}$.
 
In the first case, we consider the weighted average output state
\begin{align}\label{eq:Transformation}
    P_n\hat \rho_{n|t_c}\to P_n\hat \rho_{n} = \frac{\int_\mathcal{T} dt_c \ev{b^\dagger (t_c)b(t_c)}P_n\hat \rho_{n|t_c}}{P_S}.
\end{align}

In the latter case, we will look for photons in the signal arm in a short interval of length $T<\mathcal{T}$ around the associated click. We thus consider 
\begin{align}\label{eq:rho_trans_T}
    \hat \rho_{A|t_c}\to \hat \rho_{A|t_c,T}=\text{Tr}_{\neg T}[\hat \rho_{A|t_c}],
\end{align}
where $\text{Tr}_{\neg T}$ is the partial trace over all the signal photons outside $T$. This will lead to a zero-photon contribution due to the possibility of having no photons in the acceptance interval.

We have thus found expressions for the density matrix of our system after the heralding event only including accessible parameters of the setup. 

\subsubsection{Heralding event for the independent modes decomposition}\label{sec:bogh}
We will now consider the heralding process using the independent modes description, where the outgoing light is described by the tensor product of two mode squeezed states in Eq. \eqref{eq:TMSS}. We will assume the pump to emit pulsed light, and we will look for photons in the entire time range.

The probability of detecting a photon in the $k$'th mode in the low power limit is $P_{S,k}=\ev*{b^\dagger_k b_k}=\sinh^2(\xi_k)$, yielding the total success probability $P_{S}=\sum_\ell \sinh^2(\xi_\ell)$. The reduced density matrix of the signal field conditioned on a click at any time, can be rewritten as conditioned on a click in any of the different independent modes
\begin{align}\label{eq:rhoMIMM}
    \hat \rho \to \hat \rho_A =\frac{\sum_\ell\text{Tr}_B[b_\ell \hat \rho b_\ell^\dagger]}{P_S}.
\end{align}
We partition the density matrix as described in Eq. \eqref{eq:rhoaexp}, and find the normalized density matrix contributions and their populations up to linear order in the $P_{S,k}$. Starting with the single photon density matrix, we find  
\begin{align}\label{eq:rho1MIMM}
    \hat \rho_{1} &= \frac{\sum_{k}P_{S,k}\dyad{1_k}}{P_{S}}\\\label{eq:MIMMP1}
    P_1 &= 1-P_{S}-\frac{\sum_{k} P_{S,k}^2}{P_{S}}.
\end{align}
where $\ket{1_k}=a_k^\dagger\ket{\emptyset}$ is the state with one photon in the $k$'th signal mode and vacuum in all other modes. We can without loss of generality assume that the probabilities of detection in a specific mode are ordered in descending order, meaning $P_{S,1}\ge P_{S,2}\ge\cdots$.

We next turn to the undesired two-photon component of $\hat \rho_A$. The signal field may contain two photons in the heralded mode, or one photon in the heralded mode and one photon in one of the other modes. These possibilities are represented by the two-photon component of $\hat \rho_A$ 
\begin{align}\label{eq:rho2MIMM}
    \hat \rho_{2} &= \frac{\sum_{k}( 2P_{S,k}^2\dyad{2_k}+\sum_{\ell\neq k}P_{S,\ell}P_{S,k} \dyad{1_\ell1_k})}{\sum_j P_{S,j}^2+P_{S}^2},
\end{align}
where $\ket{1_\ell1_k}=a_\ell^\dagger\ket{1_k}$ with $\ell\neq k$. The population of the two photon state is
\begin{align}\label{eq:P2MIMM}
    P_2 & = \frac{P_{S}^2+\sum_k P_{S,k}^2}{P_{S}}.
\end{align}

Furthermore, if we allow for some loss probability $1-\eta$ in the signal arm, we have $P_{1}=\eta$ and $P_{2}= \eta^2 (P_S+\sum_k P_{S,k}^2/P_S)$ to leading order in $P_S$. By considering the ratio of the counting statistics to leading order in $P_{S}$ we can define a combination independent of the efficiency, $\eta$ in the signal arm
\begin{align}\label{eq:upsilon}
    \Upsilon\equiv\frac{P_{2}}{P_1^2 P_{S}} = 1+\sum_k \frac{P_{S,k}^2}{P_S^2},
\end{align}
where the second equality is true for this specific model.

As the right hand side depends on whether the source emits single-mode or multimode light, we can use the ratio of the counting statistics as an indicator of the multimode nature of the source. This can be exemplified by the two extreme cases: In the single-mode limit where $P_{S}=P_{S,1}$ we find $\Upsilon=2,$ and if the single-photon state is maximally mixed with $N$ equally contributing modes, we find $\Upsilon=1+1/N$. Accordingly this expression provides a rough characterisation of the multimode nature of the field, and we shall use it for this purpose below.

We have in this section, shown that if we consider a heralding click at any time, a complete description of the state of the light after the heralding process can be found through a set of parameters $P_{S,k}$ and the orthogonal modes $\ket{1_k}$. These can be found through the singular value decomposition of the Bogoliubov transformation functions detailed in Sec. \ref{sec:mode}. Alternatively, one can find them by making an eigenvalue decomposition of the single-photon density matrix contribution, found by performing the incoherent summation described in Eq. \eqref{eq:Transformation} of the expression in Eq. \eqref{eq:rho1}, where $\mathcal{T}$ is set to cover the entire time range.

\section{Characterization of the heralded states \label{sec:charchter}}

The expressions derived in the previous section allow us to describe the full density matrix of the signal field conditioned on the detection of idler photons in the few photon regime. 
However, to have a simple characterization of the source, we will consider a range of different parameters introduced below. The values of these parameters can then form the basis of further characterisation of the quality of the produced states for different protocols.

The first characterization parameters are the populations of the single-photon state and the two-photon state, $P_1$ and $P_2$ and their ratio $\Upsilon$ given in Eq. \eqref{eq:upsilon}. As the goal is to have a single photon in the signal beam after the heralding event, we want to be in a regime where $P_1\gg P_2$. In this regime, we have the second order correlation function $g^{(2)}_\text{pulse}=2P_2/P_1^2\ll 1$.

To investigate the single-photon component further, we write the density matrix in its eigenbasis
\begin{align}
    \hat \rho_1 = \sum_i w_i \dyad{\phi_i},
\end{align}
where $w_i$ is the weight associated with the eigenmode $\ket{\phi_i}$. We shall assume the weights are sorted in descending order, i.e. $w_{1}\ge w_{2}\ge \cdots$. Furthermore, we define the (mode) purity of the single-photon state as
\begin{align}\label{eq:puritym}
    \varpi_1=\Tr\qty[\hat \rho_1^2] = \sum_i w_i^2.
\end{align} 
If the single photon is emitted in a pure state, we have  $\varpi_1=1$, and for a maximally mixed state, we find $\varpi_1=1/N$, where $N$ is the number of contributing modes. The purity is crucial for any protocol relying on interference between photons from different sources. However, for some applications including Quantum Key Distribution, interference between different photons is not important, and thus the purity is less relevant. 

We now turn to the unwanted two-photon state. To make a similar characterization, we consider the decomposition of the two-photon density matrix in terms of the eigenmodes of the single-photon density matrix, $\ket{\phi_i}$. We thus define the set of populations
\begin{align}\label{eq:FIJ}
    Q_{ij} = \ev{\hat \rho_{2}}{\phi_i\phi_j},
\end{align}
where $\ket{\phi_i\phi_j}$ contains two photons in the respective modes $\ket{\phi_i}$ and $\ket{\phi_j}$.
The single-mode case is characterized by $Q_{11}=1$ as the only non-zero population.
Some protocols can mitigate errors from two-photon events \cite{jiang_fast_2007,zhao_robust_2007}; however, the effectiveness of these protocols may rely on the two photons occupying the same mode \cite{zhang_preperation_nodate}, thus requiring $Q_{ij}\approx 0$ for any $i\neq j$. Knowledge of $Q_{ij}$ can thus be used to assess the severity of errors coming from multi-photon events. 

\subsection{Hong-Ou-Mandel Visibility}
The Hong-Ou-Mandel (HOM) effect, describing how the interference of two indistinguishable photons mixed on a beamsplitter results in bunching of the two photons \cite{hong_measurement_1987}, remains one of the clearest experimental indicators of the single photon purity. The measure of the HOM effect is the HOM visibility, which we will define as \cite{branczyk_hong-ou-mandel_2017}
\begin{align}
    \mathcal{V} = 1-\frac{P_\text{cc}}{P_\text{cc,dist}},
\end{align}
where $P_\text{cc}$ ($P_\text{cc,dist}$) is the number of coincidence clicks between the two output ports in a HOM-type experiment (when the photons are completely distinguishable). If the single-photon source is perfect, i.e. the emitted light is in a pure single-photon state, the visibility is 1. However, both the impurity of the single-photon state and the non-zero two-photon component limits the maximum visibility, and the HOM visibility is thus an important tool for characterizing a single-photon source experimentally \cite{ding_-demand_2016,da_lio_pure_2022}. 

For this reason, we choose to link our characterization parameters to the HOM visibility, with the goal of indicating how a measurement of the HOM visibility can yield direct information about the multimode nature of the source. The visibility can be expanded up to linear order in $P_2$, in the regime relevant for single photon generation, where $P_1\gg P_2$. We generalise the definition given in Refs. \cite{gonzalez-ruiz_single-photon_2023,gonzalez-ruiz_preperation_nodate}
\begin{align}\label{eq:HOM}
    \mathcal{V} = \mathcal{V}_0\qty(1-F\,g^{(2)}_{\text{pulse}}),
\end{align}
where $\mathcal{V}_0$ is the single-photon visibility, i.e. the visibility when $P_2=0$ and $F$ characterizes the influence of two-photon events on the visibility. Compared to Refs. \cite{gonzalez-ruiz_single-photon_2023,gonzalez-ruiz_preperation_nodate}, which only considers $\mathcal V_0=1$, we have generalised the formula to include impure single-photon states. In doing so, we choose to take the factor $\mathcal V_0$ outside the parenthesis when defining $F$. This definition yields a better separation of effects caused by the impurity of the single-photon state and the impurity of the two-photon state into the parameters $\mathcal{V}_0$ and $F$ respectively, as is seen below.

As mentioned above, we can express $\mathcal{V}_0$ and $F$ in terms of the previously introduced characterization parameters
\begin{align}\label{eq:V0}
    \mathcal{V}_0 &= \varpi_1\\
    F &= 1+2P_1\left[1-\frac{\sum_i w_i\left(2Q_{ii}+\sum_{j\ne i} Q_{ij}\right) }{2\varpi_1}\right],\label{eq:HOMF}
\end{align}
i.e. the purity of the single-photon state is the same as the single-photon HOM visibility \cite{branczyk_hong-ou-mandel_2017}, whereas $F$ depends on the multimode character of the single-photon and two-photon states.

In the single-mode case we find $\mathcal{V}_0=F=1$. If the single-photon state remains pure and the two-photon state becomes mixed, $F$ will increase to values between $1$ and $3$ \cite{gonzalez-ruiz_single-photon_2023,gonzalez-ruiz_preperation_nodate}. If the single-photon state is in a mixed state, $\mathcal{V}_0$ will decrease as the purity decreases. In the limit where the single-photon state is maximally mixed, i.e. the single-photon state populates each eigenmode equally, $F$ becomes 1. In the opposite limit where the single-photon state mostly populates a single eigenmode and the two-photon state mostly populates a single combination of eigenmodes, we find $F$ to decrease slightly from the corresponding value when $\mathcal{V}_0=1$. Thus, by extracting $\mathcal{V}_0$ and $F$ from a HOM-type experiment, one can gain information about the multimode nature of the emitted light.

\begin{figure}
    \centering
    \includegraphics[width=\columnwidth]{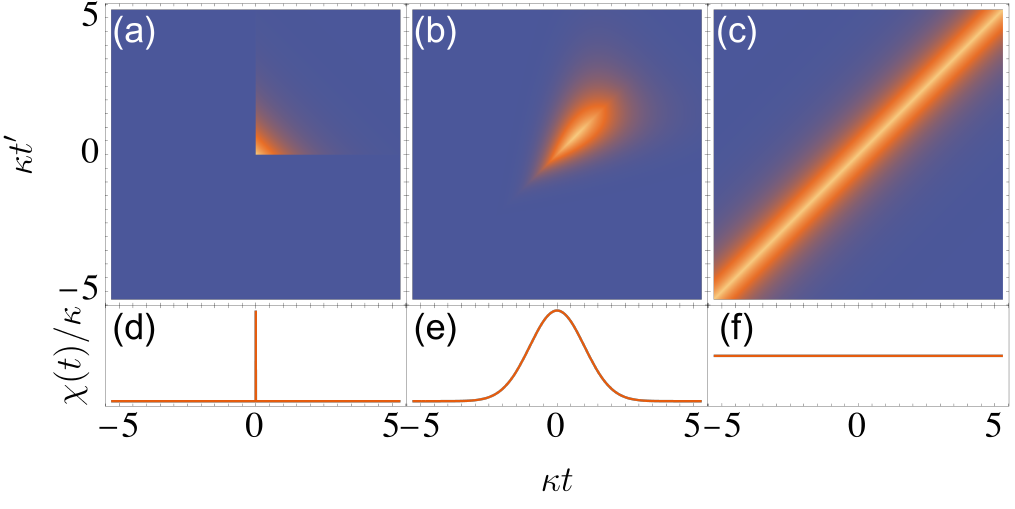}
    \caption{
    \textbf{(a)-(c)} The correlations between a photon in the signal beam at time $t$ and a photon in the idler beam at time $t'$ in the weak driving limit. The highest (lowest) value is attained in the yellow (blue) regions. \textbf{(d)-(f)} The temporal shape of the interaction $\chi(t)$. \textbf{(a), (d)} Infinitely short pulse. \textbf{(b), (e)} Gaussian pulse with $\kappa\sigma=1$. \textbf{(c), (f)} CW laser.}
    \label{fig:G2_cases}
\end{figure}
\section{Quantitative results for different driving methods  \label{sec:Results}}
In this section, we examine three different temporal profiles of the driving field: An infinitely short pulse, a finite duration Gaussian pulse and a continuous wave laser. See Fig. \ref{fig:G2_cases}, for the pulse shapes and the lowest order correlations between photons in the signal and idler beams for the three cases. We will characterize the quantum state heralded in the three cases and make use of the parameters introduced in the previous section to interpret and explain the results.

\subsection{Short pulse\label{ssec:SP}}

We first consider an infinitely short pump pulse, meaning $\chi(t)=x \delta(t)$, where $x$ is the dimensionless strength of the interaction. We sketch the correlations between photons in the two beams in Fig. \ref{fig:G2_cases}(a). We see that the light in the two beams are temporally uncorrelated matching the experimental result in Ref. \cite{zhang_preparation_2011}, where the output states in the signal and idler beams is shown to have uncorrelated frequencies if a narrow pump pulse is employed. 

The infinitely short pulse first creates a number of excitations of the intra-cavity modes, and these excitations subsequently leak into uncorrelated exponentially decaying single-mode wave packets in the signal and idler beams. Hence, the full density matrix before the heralding event is described by
\begin{align}
    \hat \rho = \sum_n A_{n} (a^\dagger_{\phi_\text{SP}})^n(b^\dagger_{\phi_\text{SP}})^n \dyad{\emptyset}(a_{\phi_\text{SP}})^n(b_{\phi_\text{SP}})^n,
\end{align}
where $A_n$ is the weight of the $n$'th component, which can be calculated from the two point correlation functions given below, and $a^\dagger_{\phi_\text{SP}}$ ($b^\dagger_{\phi_\text{SP}}$) is the creation operator of a photon in the signal (idler) field in the wave packet shaped by $\phi_\text{SP}(t)$, i.e.
\begin{align} 
    a^\dagger_{\phi_\text{SP}} = \int_{-\infty}^\infty \phi_\text{SP}(t) a(t),
\end{align}
where $\phi_\text{SP}(t)$ is defined as
\begin{align} 
    \phi_\text{SP}(t) = \sqrt{\kappa}\, e^{-\kappa t /2} \Theta(t).
\end{align}

We obtain the two-point correlation functions of the signal and idler fields from the full density matrix to fourth order in $x$
\begin{subequations}\label{eq:SP_expect}
    \begin{align}
        \ev*{ a^\dagger(t) a(t')}&= \phi_\text{SP}(t)\phi_\text{SP}(t')\left(x^2+\frac{x^4}{2}\right)\\
        \ev*{a(t) b(t')} &=-i  \phi_\text{SP}(t)\phi_\text{SP}(t')\left(x+\frac{3x^3}{4}\right).
    \end{align}
\end{subequations}
From the correlation functions we find the success probability to lowest order in $x$ as $P_S=x^2$. 

The heralding event only influences the state in the signal beam by changing the populations of the different $n$-photon density matrix components, meaning the normalized $n$-photon density matrix is $\hat \rho_n = (a^\dagger_{\phi_\text{SP}})^n\dyad{\emptyset}(a_{\phi_\text{SP}})^n/n!$. Using Eq. \eqref{eq:rhos}, we find the populations of the conditional single-photon state and the conditional two-photon state to second order in $x$ to be $P_1=1-2x^2$ and $P_2=2x^2$. As the emitted light is single mode, the HOM-visibility parameters are easily found to be $\mathcal{V}_0=F=1$, and the ratio of the counting statistics is found to be $\Upsilon=2$ in the low driving regime.

\begin{figure*}
    \centering
    \includegraphics[width=\textwidth]{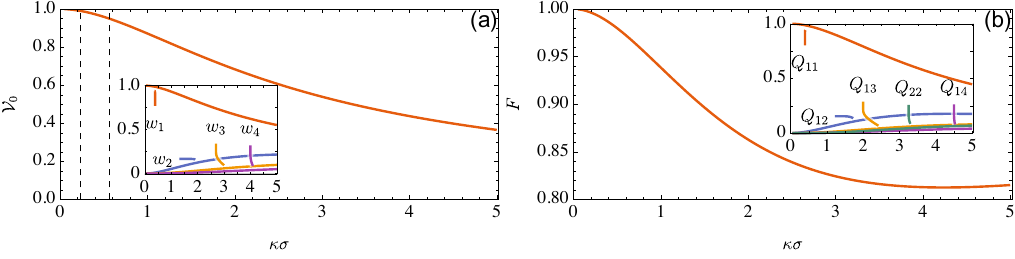}
    \caption{
    Characterization of a heralded SPDC source driven by a Gaussian pump pulse of width $\sigma$.
    \textbf{(a)} Purity of the single-photon density matrix for weak driving. The two dashed lines mark the two values of $\sigma$ for which $\mathcal{V}_0=0.99$ and $\mathcal{V}_0=0.95$. The four largest weights of the single photon eigen-decomposition are plotted in the inset.
    \textbf{(b)} Linear response of HOM visibility to two-photon events, see Eq. \eqref{eq:HOM}. In the inset, we show the populations of different combinations of single-photon eigenmodes in the two-photon state.
    }
    \label{fig:num}
\end{figure*}

\subsection{Gaussian pump pulse\label{ssec:pulse}}
We next investigate driving of the SPDC source with a finite duration pump pulse. We assume that the pump strength takes the form $\chi(t)= (x/\sqrt{2\pi\sigma^2}) \exp(-t^2/2\sigma^2)$, where $\sigma$ defines the width of the pulse and $x$ is again a dimensionless strength. We show the intensity correlations between the two beams for $\kappa \sigma=1$ in Fig. \ref{fig:G2_cases}(b). The photons in the two beams now have some temporal correlations. To which degree photons in the two beams are temporally correlated is determined by the width of the pulse.

In the heralding protocol, we will not condition on a specific detection time of the idler photon; and we will hence average over all possible detection times. 
This means that results derived in Section \ref{sec:bogh} are applicable here. We will follow the method introduced earlier. We find the single photon density matrix contribution by taking the trace over the possible click times of the expression in Eq. \eqref{eq:rho1}. This density matrix is then diagonalized to find the weights of the eigenmodes. This allows us to compute the other density matrix contributions, and thus the characterization parameters. One can alternatively use the general density matrix in Eq. \eqref{eq:rhos} and express the characterization parameters directly in terms of integrals. These integrals are provided in Appendix \ref{sec:AppNum} and can be solved numerically. 

In the inset of Fig. \ref{fig:num}(a) we plot the weights of the most prominent eigenstates of the single photon density matrix for different pulse widths. When $\kappa\sigma \ll 1$ the single-photon state is almost exclusively populated by one photon in the same state, i.e. in this regime the light behaves close to the single-mode limit. When $\sigma$ increases the population of this mode decreases, and the other eigenmodes become more prominent, necessitating the full multimode description. This is also clear when considering the single-photon purity, or equivalently the single-photon HOM visibility, plotted in Fig. \ref{fig:num}(a). The purity is close to unity in the narrow pulse regime, but decreases when transitioning to longer pulses. From Eq. \eqref{eq:MIMMP1} we can relate the single photon population and the single photon purity through $P_1 = 1-(1+\varpi_1)P_S$.

We next turn to the two-photon component. The population of the two-photon density matrix component can be found from Eq. \eqref{eq:P2MIMM}, yielding to leading order in $P_S$
\begin{align}
    P_2 = P_S(1+\varpi_1),
\end{align}
corresponding to the ratio of counting statistics $\Upsilon=1+\varpi_1$.
In the single mode case, where $\varpi_1=1$, we obtain $P_2=2P_{S}$. In the opposite limit, where the single-photon state is maximally mixed $P_{2} =(1+1/N) P_{S}$. This relation reveals a trade-off between the purity of the single-photon state and the population of the unwanted two-photon state. A high single-photon purity yields a relatively higher probability of emitting multiple photons reflecting photon bunching in a single mode squeezed state. However, if the coherence of the light from different sources is less important, it may be an advantage to have a less pure single-photon state, to gain a smaller population of the two-photon state.
As described above, the mode purity of the single-photon state is close to unity in the short pulse regime; however, as $\sigma$ increases the single-photon purity decreases. The two-photon population will follow this behaviour and decrease as the pump pulse is made wider.

We can express the populations of the two-photon state, defined in Eq. \eqref{eq:FIJ}, by considering the two-photon density matrix for this case in terms of the weights of the single-photon density matrix
\begin{align}
    Q_{ij} = \frac{2w_iw_j}{1+\varpi_1},
\end{align}
which tells us that the two photon state is pure, if the single-photon state is pure, and that the two photon state is maximally mixed, if the single photon state is maximally mixed.

The largest two-photon populations are plotted in the inset of Fig. \ref{fig:num}(b). When $\kappa\sigma \ll 1$ the two-photon state is almost exclusively populated by two photons in the same state. When $\sigma$ increases the population of this state decreases, and other combinations of eigenmodes become more prominent. This result corroborates the conclusion from the characterization of the single-photon component: a single-mode description is a good approximation in the narrow pulse regime, but when the pulse becomes wider, a multimode description is necessary to account for the quantum state.

We now consider the linear response of the HOM visibility on two photon events, and find it in terms of the single-photon weights
\begin{align}\label{eq:MIMMHOMF}
    F = 1+\frac{2P_1}{1+\varpi_1}\left(\varpi_1-\frac{ \sum_i w_i^3}{\varpi_1}\right).
\end{align}
If the single-photon state is pure or maximally mixed, the value of $F$ becomes 1, matching the calculations from before. However, between these two extremes we find $F$ to be smaller than 1.

The parameter $F$ is plotted in Fig. \ref{fig:num}(b). $F$ decreases initially when the pulse is made broader, going from the single-mode value of 1 when $\kappa \sigma \ll 1$ to $\sim 0.81$ for $\kappa \sigma \sim 4$, with a slight increase afterwards. The decrease in $F$ means that the HOM visibility is less influenced by the two-photon component compared to the single mode case. The reduction in $F$ is intimately linked to the decrease in single-photon visibility from the impure single-photon component. For a mixed single-photon state, the additional photon might have an overlap with the multiple components from the impure single-photon state, thus the second photon is less disruptive for the HOM visibility than in the single mode case. However, since we are operating in the limit of single-photon generation where $g^{(2)}_\text{pulse}\ll 1$, the advantage gained from a reduction in $F$ cannot mitigate the decrease in HOM visibility from the impure single-photon states.

\begin{figure*}
    \includegraphics[width=\textwidth]{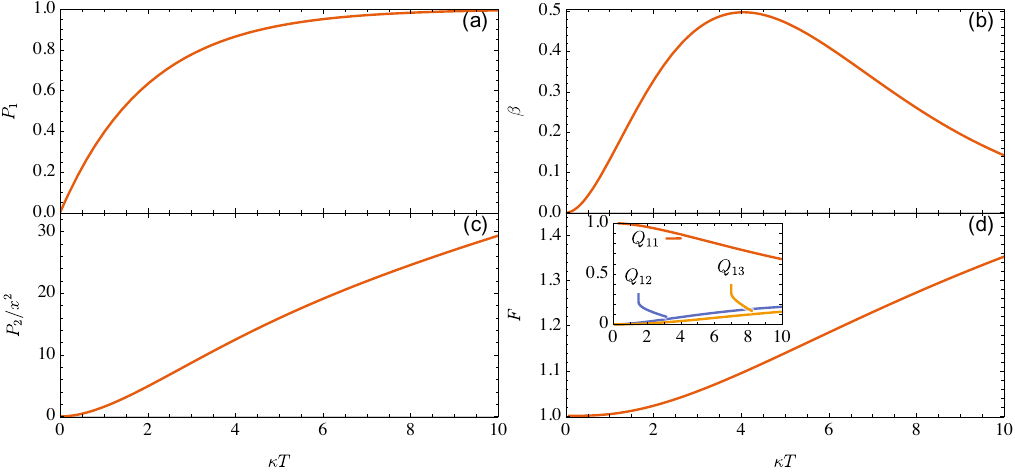}
    \caption{Characterization of a SPDC source driven with a constant pump field and accepting clicks within the acceptance time interval $T$ around the detector click.
    \textbf{(a)} Population of the single-photon state in the weak driving limit.
    \textbf{(b)} Lowest order correction to the purity of the single-photon density matrix, $1-\varpi_1=\beta x^2$. 
    \textbf{(c)} Lowest order contribution to the population of the two-photon density matrix component. 
    \textbf{(d)} The linear response of HOM to two-photon events, see Eq. \eqref{eq:HOM}.
    The inset shows the populations of the expansion of the two-photon state on eigenmodes of the single-photon density matrix. The most prominent eigenmode is $\ket{\phi_\text{CW}}$ given in Eq. \eqref{eq:phi_cw}. The other eigenmodes are found numerically using the value $x=\sqrt{10^{-3}/2\kappa T}$.}
    \label{fig:CW}
\end{figure*}

\subsection{Continuous wave\label{ssec:CW}}
We now turn to the continuous wave setting with a constant pump laser, setting $\chi(t)=x \kappa$, such that $x$ is dimensionless. The correlations between photons in the two arms for this setting are shown in Fig \ref{fig:G2_cases}(c). The signal and idler beams are highly temporally correlated under these experimental conditions. As we are working in a continuous setting, the probability of success increases linearly with the time $\mathcal{T}$. The relevant quantity is thus the success rate $P_S/\mathcal{T}$. Furthermore, this effectively partitions the long time $\mathcal{T}$ into smaller sections, creating temporal multiplexing, where we try to obtain the heralding click in each section. This is a great advantage, when the success probability is small, e.g. in the establishment of long-distance quantum network, as the success probability is effectively multiplied by the number of multiplexing modes.

To evaluate the quality of the state, we find the two point correlations up to the appropriate order in $x$ 

    \begin{subequations}
    \begin{widetext} 
        \begin{align}\label{eq:evsCW}
            \ev{a^\dagger(t) a(t')}&=  e^{-\frac{\kappa}{2}\abs{t-t'}}x^2\kappa\pqty{2+\kappa\abs{t-t'}} + \order{x^3}\\\label{eq:evsCW2}
            \ev{b^\dagger(t_c) b(t_c)}&= 2x^2\kappa\bqty{1+4x^2} +\order{x^5}\\
            \ev*{a(t)b(t')}&= -ie^{-\frac{\kappa}{2}\abs{t-t'}}x\kappa\qty{1+x^2\bqty{4+\kappa\abs{t-t'}\pqty{2+\frac{\kappa\abs{t-t'}}{2}}}} + \order{x^5}.
        \end{align}
\end{widetext}
\end{subequations}

We make the transformation of the density matrix described in Eq. \eqref{eq:rho_trans_T}, where we accept photons in the signal arm inside a time interval of length $T$ selected symmetrically around the heralding click. While we omit the subscript $T$ in the text, all characterization parameters depend on the length of the acceptance interval. We use the results from Sec. \ref{sec:model} to extract analytical expressions for the density matrix components, $P_1$, $P_2$, $\varpi_1$ and $Q_{11}$ from the two point correlation functions. These expressions are given in Appendix~\ref{sec:AppCW}. 

The success rate $P_S/\mathcal{T}$ can be calculated from Eqs. \eqref{eq:PS} and \eqref{eq:evsCW2}, to give $P_S/\mathcal{T}=2\kappa x^2+\order{x^4}$. We will thus expand all the quantities to second order in $x$. 

The conditional single-photon population $P_1$ is plotted in Fig. \ref{fig:CW}(a). The average time between the emission of the signal and idler photons is set by the cavity decay time $\sim 1/\kappa$, hence, for short (long) time intervals $\kappa T\ll 1$ ($\kappa T\gg 1$), the probability of having a photon inside the acceptance interval $T$ is small (approaches unity). This behaviour is seen in the population of the single-photon state.

We next consider the purity of the single-photon state. By examining Eq.~\eqref{eq:rho1cw}, we note that $\rho_1(t_1,t_2)$ can be expressed as a product of the most populated mode function evaluated at $t_1$ and $t_2$, respectively, in the weak driving limit. Thus, the density matrix can be expressed as  $\hat \rho_1=\dyad{\phi_\text{CW}}$, where 
\begin{align}\label{eq:phi_cw}
    \ket{\phi_\text{CW}} = \sqrt{\frac{\kappa}{2(1-e^{-\frac{\kappa T}{2}})}} \int_{t_c-\frac{T}{2}}^{t_c+\frac{T}{2}} dt \, e^{-\frac\kappa 2 \abs{t-t_c}} \ket{t}.
\end{align}
However, the single photon density matrix can become mixed to lowest order in $\chi(t)$ if the setup is changed to include limitations from real devices. In Ref.  \cite{sonoyama_analysis_2022} the finite temporal resolution of the detectors leads to a mixed single photon density matrix.

When $x$ increases, multiple pairs of photons can be emitted making it possible that the photon associated with the heralding click is outside the acceptance interval, while a signal photon from a different pair is inside the acceptance interval. This, leads to a decrease in the purity of the single-photon state. The most populated mode continues to be $\ket{\phi_\text{CW}}$, and the other eigenmodes can be found numerically. We find that the leading contribution to $1-\varpi_1$ is of order $\order{x^2}$; thus, by introducing a parameter $\beta$ we can write the purity as $\varpi_1 = 1-\beta x^2 + \order{x^4}$, where $\beta$ is plotted for a range of time intervals in Fig. \ref{fig:CW}(b).

We now characterize the two-photon state, starting with the second order contribution of $P_2$, which is plotted in Fig. \ref{fig:CW}(c). As $T$ becomes large, the two-photon population increases linearly with longer time intervals, which stems from an increased contribution from uncorrelated photons with a constant photon flux. Thus, while the conditional population of the single-photon state stagnates, the error rate from two-photon events increases.

We now move to the $F$-factor in the HOM visibility and the populations of the two-photon state $Q_{ij}$ plotted in Fig. \ref{fig:CW}(d). For short time intervals, the two-photon state is solely populated by two photons in $\ket{\phi_\text{CW}}$ matching the single-mode description, which means the $F$ factor is close to unity in this regime. For longer acceptance intervals, combinations where one photon is in a different mode than $\ket{\phi_\text{CW}}$ get increasingly populated. As we begin to add more modes for the second photon to occupy in the two-photon state, $F$ increases.

As expected, our characterization highlights a compromise when designing experiments based on these sources. For short acceptance intervals, the error rate from two-photon events is negligible, however, the probability of getting any photon after the heralding event is small, as the signal photon accompanying the detected photon might be outside the acceptance interval. Conversely, choosing a long interval length $T$, ensures that the signal photon is inside the interval, but the error rate from two-photon events increases, due to the contamination by photons generated at other times.
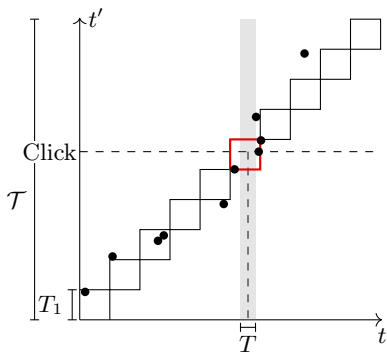
\begin{figure}
    \centering
    \begin{tikzpicture}[scale=0.2,
            dots/.style={circle,fill=black,inner sep=0.4mm}]
        \draw[->] (0,0) -- (20.1,0) node[below, pos=1] {$t$};
        \draw[->] (0,0) -- (0,20.1) node[right, pos=1] {$t'$};
        \draw (2,0) -- (2,2) -- (0,2);
        \foreach \x in {1,2,3,4,5,6,7,8,9}{
            \draw (2*\x+2,2*\x+2) rectangle (2*\x,2*\x);

        };

        \draw (-0.2,0) -- (-0.8,0) (-0.2,2) -- (-0.8,2);
        \draw (-0.5,0) -- (-0.5,2) node [left, pos=0.5] {$T_1$};
        \draw (-3.3,0) -- (-2.7,0) (-3.3,20) -- (-2.7,20);
        \draw (-3,0) -- (-3,20) node [left, pos=0.4] {$\mathcal T$};

        \pgfmathsetmacro{\T}{1}
        \pgfmathsetmacro{\tc}{11.186}
        \draw[dashed] (0,\tc) -- (20,\tc) (\tc,\tc) -- (\tc,0);
        \draw (\tc-\T/2,-0.2) -- (\tc-\T/2,-0.8) (\tc+\T/2,-0.2) -- (\tc+\T/2,-0.8);
        \draw (\tc-\T/2,-0.5) -- (\tc+\T/2,-0.5) node [below, pos=0.5] {$T$};
        \fill[fill=gray, fill opacity=0.2] (\tc-\T/2,0) rectangle (\tc+\T/2,20);
        \draw[red!90!black,thick] (10+2,10+2) rectangle (10,10);
        \node at (-2,\tc) [fill=white,inner xsep=0pt,inner ysep=1pt] {Click};

        \node[dots] at (0.36 ,  1.856) {};
        \node[dots] at ( 2.182,  4.214) {};
        \node[dots] at ( 5.202,  5.256) {};
        \node[dots] at ( 5.588,  5.614) {};
        \node[dots] at ( 9.574,  7.702) {};
        \node[dots] at (10.294, 10.01 ) {};
        \node[dots] at (11.9  , 11.186) {};
        \node[dots] at (12.05 , 11.934) {};
        \node[dots] at (11.724, 13.498) {};
        \node[dots] at (14.944, 17.716) {};
    \end{tikzpicture}    
    \caption{Illustration of the multiple independent modes model used in the CW case. Each point represents a photon pair, present in the signal beam at time $t$ and in the idler beam at time $t'$.
    We have indicated the length of a time bin, $T_1$ and the acceptance interval, $T$, on the $t$-axis, and we have indicated the total duration, $\mathcal{T}$ on the $t'$ axis. 
    When a photon is detected in the idler beam, it heralds the presence of a signal photon in the area highlighted by the gray rectangle on the $t$-axis. Furthermore, in the independent modes picture, the presence of the photon in the mode highlighted in red is heralded by the detection. The actual temporal distribution of photon pairs leaves the possibility of no (multiple) photons in the gray interval, in particular if $T$ is chosen to be small (large). Similar arguments can be made for the $T_1$.
    The photon pairs are sampled from the distribution shown in Fig. \ref{fig:G2_cases}(c), with $T$ and $T_1$ set to $1/\kappa$ and $2/\kappa$ respectively. 
    }
    \label{fig:CW_MIMM}
\end{figure}

\subsubsection{Description by a multiple independent modes model}
As the heralding scheme in this case is based on the detection at a specific time, we cannot apply the theory developed in Section \ref{sec:bogh}. However, to guide intuition when setting up an experiment based on a CW driven SPDC source, we now introduce a simplified model to highlight the trade-offs between different properties of the model inherent from the full multimode nature. Since sources are for simplicity typically described in terms of single mode models, we shall here develop a model, where the source is described as a collection of independent single modes with modified parameters.

\begin{figure*}
    \includegraphics[width=\textwidth]{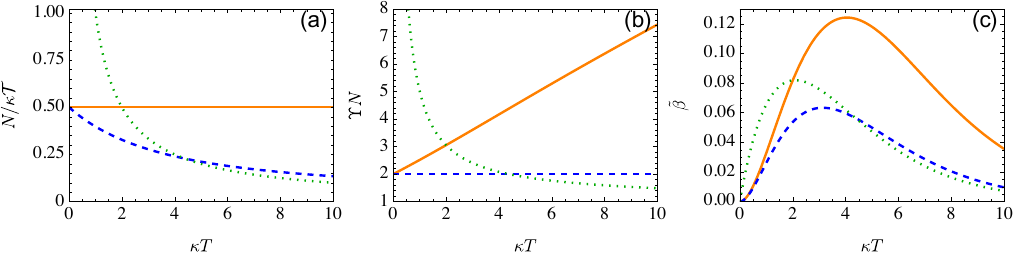}
    \caption{Characterization of the different approximate models introduced in the text. Model (i), (ii) and (iii) are shown with the solid orange, dashed blue and dotted green curves respectively. 
    \textbf{(a)} The number of time bin modes pr. total time.
    \textbf{(b)} The ratio between single- and two-photon counting statistics.
    \textbf{(c)} The parameter $\tilde \beta$ which represents the correction to the purity of the single-photon density matrix to linear order in $P_{S,1}$, $\tilde\beta P_{S,1} = 1-\varpi_1$.}
    \label{fig:g2sm}
\end{figure*}

We consider the model sketched in Fig. \ref{fig:CW_MIMM}. The total duration, where we look for photons, $\mathcal T$, is partitioned into $N$ uncorrelated time bin modes of length $T_1=\mathcal T/N$. The probability for getting a heralding click in any particular time bin mode is $P_{S,1}=2x^2\kappa T_1+\order{x^4}$. 

The properties of the model, e.g. the number of modes and the purity of the single-photon state, depends on the relation between $T_1$ and the acceptance interval $T$. Hence, by changing the relation between $T_1$ and $T$, we optimize the model for different properties. We consider three separate and equally valid models, based on different choices:
\begin{enumerate}[(i)]
    \item The number of modes per time interval is defined by the cavity decay rate. Here, $T_1$ is independent of $T$.
    \item The ratio of the counting statistics follows the single mode prediction. The relation between $T_1$ and $T$ is chosen to ensure this.
    \item The length of a mode is set by $T$, i.e. $T_1=T$.
\end{enumerate}

To compare the different models, we focus on three parameters, which impact protocols relying on a single-photon source. The first parameter is the number of modes pr. time, which yields the number of modes available for multiplexing. The second parameter is the ratio of the counting statistics, as expressed through $\Upsilon$, yielding a measure of the relative occurrence of two-photon events to single-photon events. As the total success probability grows with the total time $\mathcal{T}$, $\Upsilon$ decreases with the total time, since it is normalised to the overall success probability. To remove the dependence on the total time, we will consider $\Upsilon N$, which is the counting statistics with respect to a single time bin mode. This is found to be
\begin{align}
    \Upsilon N = \frac{P_{2}/x^2}{P_{1}^2}\frac{1}{2 \kappa T_1},
\end{align}
and is equal to 2 in the single-mode case.
The third parameter is the correction to the purity of the single-photon state to linear order in $P_{S,1}$, defining $1-\varpi_1 = \tilde \beta P_{S,1} + \order{P_{S,1}^2}$. These properties are plotted for the three models in Fig. \ref{fig:g2sm}, and the precise definitions of the three models are given below.

We see a large variation between the models for the first two parameters; however, for the correction to the purity, even for a relatively large success probability for any particular mode of $P_{S,1}=0.15$ the purity of the single-photon state is bounded by $\varpi_1\gtrsim 0.98$ for all models. We will thus focus on the other two parameters in the comparison.

In model (i), the modes have a fixed duration. In the limit where the acceptance interval is very narrow, i.e. when $\kappa T\ll 1$, we expect the emitted light to be single-mode, meaning the ratio of counting statistics, as expressed through $\Upsilon N$, should approach the single-mode value. Solving for $T_1$ in this case, we find $T_1=2/\kappa$, yielding the duration of the time bins for this model. We now consider the ratio of counting statistics, $\Upsilon N$. For a fixed time bin duration, we can consider the populations of the single- and two-photon states directly from Figs. \ref{fig:CW}(a) and \ref{fig:CW}(c). We see that $P_1$ increases asymptotically towards unity for long time intervals, whereas $P_2/x^2$ is not bounded in the weak driving limit. Thus, the ratio of counting statistics will increase as the acceptance interval increases, as is seen in Fig. \ref{fig:g2sm}(b). This tells us that the relative two-photon error increases with the acceptance interval if we fix the length of the time bins.

We now turn to model (ii). In this model, $T_1$ is chosen to ensure that the counting statistics obey the single-mode relation $\Upsilon N = 2$. In this case, the errors arising from two-photon events are fixed at the single mode predicted level. This is compensated by changing the number of modes available for multiplexing. Solving for the number of modes, we find 
\begin{align}
    N=\frac{4P_{1}^2}{P_{2}}x^2\kappa\mathcal T,
\end{align}
which depends on the length of the acceptance interval through Eq. \eqref{eq:probscw}. In Fig. \ref{fig:g2sm}(a), we see that the number of modes decrease when increasing $T$ for this model. Thus, the price of the constant error rate from two-photon events is the number of modes available for multiplexing when increasing the acceptance interval.

We finally consider model (iii). In this model, we fix $T_1$ to the length of the interval, i.e. $T_1=T$. Hence, the number of modes available for multiplexing is inversely proportional to $T$. When $\kappa T$ is small, many modes are available for multiplexing. The relative error rate from two-photon events is however large. We find the opposite case, when $\kappa T$ is large.

From the three approximate and equally valid models, we see that the properties cannot be determined independently. The problem is that the emitted light is fundamentally multimode and the modes are not independent. Thus, when constructing this simplified model where the modes are assumed to be independent, we have to compensate at some point. If one chooses a relatively large number of modes, the error rate from two-photon events is larger than the predicted value in a single-mode description. Alternatively, if the single-mode predicted two-photon error is assumed, fewer modes are available for multiplexing. The rate of two-photon errors and the number of multiplexing modes only reach their desired values simultaneously in the regime $\kappa T\ll 1$, where the conditional probability of having any photons inside the acceptance interval is small cf. Fig. \ref{fig:CW}(a).

\section{Conclusion\label{sec:Conclusion}}
We have developed a full multimode description to characterize the light from an SPDC source under three different scenarios, where the OPO is driven using an infinitely short pulse, the OPO is driven using a Gaussian pulse of varying width, or the OPO is driven continuously. In the first case, we show the light follows the single-mode behaviour. In the second case, the light can be described as multiple independent modes. As the width of the pulse increases a larger number of modes influence the state of the light. For a fixed total success probability, the population of the two-photon component decreases as the light becomes more multimode at the expense of the purity of the single-photon state. In the CW case, the state is assigned to a mode or modes centered around the click time. Again the multimode description is necessary to explain the physics of the system. Nevertheless, the behavior in this scenario can be approximated by defining simplified models with independent modes capturing the essential effects by adjusting the number of modes and counting statistics. Our findings can thus guide the design of various quantum information protocols to achieve the best possible outputs when employing heralded single-photon sources.

\begin{acknowledgments}
The authors acknowledge the support of Danmarks Grundforskningsfond (DNRF 139, Hy-Q Center for Hybrid Quantum Networks). This work was supported by the project QIA-Phase 1 which has received funding from the European Union’s Horizon Europe research and innova- tion programme under grant agreement No. 101102140

\end{acknowledgments}

\bibliography{references}
\newpage

    \appendix
\section{Singular value decomposition of a general two photon Bogoliubov transformation matrix}\label{sec:AppBog}
We will in this appendix derive the results for a singular value decomposition of the Bogoliubov transformation functions. We will start by descretizing the time axis, meaning for all of the fields $d(t_i)$ are written as $d_i$. The most general Bogoliubov transformation is
\begin{align}
    \begin{pmatrix}
        a_i\\b^\dagger_i
    \end{pmatrix} = \sum_j\begin{pmatrix}
        u^{a}_{i,j}&v^{ab}_{i,j}\\ (v^{ba}_{i,j})^*&(u^{b}_{i,j})^*
    \end{pmatrix}\begin{pmatrix}
        a_{\text{in},j}\\b^\dagger_{\text{in},j}
    \end{pmatrix}.
\end{align}
We write the different $u^k_{i,j}$ and $v^{k\bar{k}}_{i,j}$, where $\bar{k}$ is $b$ ($a$) when $k$ is $a$ ($b$), as matrices $\textbf{U}_k$ and $\textbf{V}_{k\bar{k}}$. From the commutation relations we obtain the identities
\begin{subequations}
    \begin{align}\label{eq:bogrel1}
        I &= \textbf{U}_k\textbf{U}_k^\dagger-\textbf{V}_{k\bar{k}}\textbf{V}_{k\bar{k}}^\dagger = \textbf{U}_k^\dagger\textbf{U}_k-\textbf{V}_{\bar{k} k}^T\textbf{V}_{\bar{k} k}^*\\\label{eq:bogrel2}
        0 &= \textbf{U}_a\textbf{V}_{ba}^T-\textbf{V}_{ab}\textbf{U}_b^T = \textbf{V}_{ba}^T\textbf{U}_b^*-  \textbf{U}_a^\dagger\textbf{V}_{ab},
    \end{align}
\end{subequations}
where $I$ is the the identity matrix. 
From the first expression, we see that $\textbf{U}_k\textbf{U}_k^\dagger$ ($\textbf{U}_k^\dagger\textbf{U}_k$) and $\textbf{V}_{k\bar{k}}\textbf{V}_{k\bar{k}}^\dagger$ ($\textbf{V}_{\bar{k} k}^T\textbf{V}_{\bar{k} k}^*$) commute, thus they are diagonalized by the same basis. From diagonalizing the matrix $AA^\dagger$ ($A^\dagger A$) one obtains the left (right) singular vectors of the singular value decomposition of $A$. We thus note that $\vb{U}_k$ and $\vb{V}_{k\bar{k}}$ ($\vb{V}_{\bar{k}k}^*$) has the same left (right) singular vector. From Eq. \eqref{eq:bogrel2} we note that $\vb{U}_a$ and $\vb{U}_b$ ($\textbf{V}_{ab}$ and $\textbf{V}_{ba}$) must have the same singular values.
We make a singular value decomposition of $\textbf{U}_a$, $\textbf{U}_b$, $\textbf{V}_{ab}$ and $\textbf{V}_{ba}$, using the knowledge just attained

\begin{subequations}
    \begin{align}
        \vb{U}_a &= \sum_\ell \lambda_{\ell} \vb{f}_\ell \vb{g}_\ell^T\\ 
        \vb{U}_b &= \sum_\ell \lambda_{\ell} \vb{p}_\ell \vb{q}_\ell^T\\ 
        \vb{V}_{ab} &= \sum_\ell \mu_{\ell} \vb{f}_\ell \vb{q}_\ell^\dagger\\
        \vb{V}_{ba} &= \sum_\ell \mu_{\ell} \vb{p}_\ell \vb{g}_\ell^\dagger,
    \end{align}
\end{subequations}
where $\lambda_\ell$ and $\mu_\ell$ are singular values, $\vb{f}_\ell$ and $\vb{p}_\ell$ are left singular vectors and $\vb{g}_\ell^T$ and $\vb{q}_\ell^\dagger$ are right singular vectors.
Putting this back into the Bogoliubov transformation we find

\begin{align}
    \begin{pmatrix}
        a_i\\b^\dagger_i
    \end{pmatrix} = \sum_{j,\ell}\begin{pmatrix}
        \lambda_{\ell}[\vb{f}_\ell \vb{g}_\ell^T]_{i,j} &\mu_{\ell} [\vb{f}_\ell \vb{q}_\ell^\dagger]_{i,j}\\ \mu_{\ell} [\vb{p}_\ell^*\vb{g}_\ell^T]_{i,j} & \lambda_{\ell} [\vb{p}_\ell^*\vb{q}_\ell^\dagger]_{i,j}
    \end{pmatrix}\begin{pmatrix}
        a_{\text{in},j}\\b^\dagger_{\text{in},j}
    \end{pmatrix},
\end{align}
where $A_{i,j}$ is the matrix entry in the $i$'th row and the $j$'th column.
We now make a basis transformation, transforming the output (input) fields $a_i$ and $b_i$ ($a_{\text{in},j}$ and $b_{\text{in},j}$) into the basis defined by the vectors $\vb{f}_\ell$ and $\vb{p}_\ell$ ($\vb g_\ell^T$ and $\vb q_\ell^T$) respectively. Finally, the singular values must fulfill $\lambda_\ell^2-\mu_\ell^2=1$, leading us to write $\lambda_\ell=\cosh(\xi_\ell)$ and $\mu_{\ell}=\sinh(\xi_\ell)$. We thus obtain

\begin{align}
    \begin{pmatrix}
        a_\ell\\b^\dagger_\ell
    \end{pmatrix} = \begin{pmatrix}
        \cosh(\xi_{\ell}) & \sinh(\xi_{\ell}) \\ \sinh(\xi_{\ell})  & \cosh(\xi_{\ell})
    \end{pmatrix}\begin{pmatrix}
        a_{\text{in},\ell}\\b^\dagger_{\text{in},\ell}
    \end{pmatrix}.
\end{align}
We can now make the same transformations into the Schrödinger picture as in the main text. The primary difference between this scenario and the one considered in the main text, is that the mode shape of the signal and idler fields are not identical.

    \section{General Expression for density matrices \label{sec:App2}}
    In this appendix, we will show how the general form for the density matrices in Eq.~\eqref{eq:dmgen} is found. We start from Eq.~\eqref{eq:c2n}, expand the density matrix in a sum of components with $m$ photons \eqref{eq:rhoaexp} and explicitly write out the trace, yielding
    \begin{widetext}
    \begin{align}
        C_{2n}(\{t\}) = \sum_{m=0}^\infty\sum_{k=0}^{\infty}\frac{1}{k!}\int  d\tau_1 \cdots d\tau_k \mel{\tau_1,\dots,\tau_k,t_1,\dots,t_n}{P_m\hat \rho_{m}}{t_{n+1},\dots,t_{2n},\tau_k,\dots,\tau_1}.
    \end{align}    
    The matrix element is zero when $m\neq n+k$, we thus set $m=n+k$. We now introduce $N$, as the maximal number of photons we will include, meaning $\hat \rho_n=0$ for $n>N$. We then find
    \begin{align}\label{eq:proff_start}
        C_{2(N-n)}(\{t\}) = &\,\sum_{k=0}^{n}\frac{1}{k!}\int  d\tau_1\cdots d\tau_k\mel{\tau_1,\dots,\tau_k,t_1,\dots,t_{N-n}}{P_{N-n+k}\hat \rho_{N-n+k}}{t_{N-n+1},\dots,t_{2(N-n)},\tau_k,\dots,\tau_1},
    \end{align}    
    We want to invert this equation, to find $\mel{t_1,\dots,t_{N-n}}{\hat \rho_{N-n}}{t_{N-n+1},\dots,t_{2(N-n)}}$. We postulate that
    \begin{align}\label{eq:post}
        P_{N-n}\mel{t_1,\dots,t_{N-n}}{\hat \rho_{N-n}}{t_{N-n+1},\dots,t_{2(N-n)}} = \sum_{k=0}^{n}\frac{(-1)^k}{k!} \int  d\tau_1\cdots d\tau_k C_{2(N-n+k)}(\{t\}'),
    \end{align}    
    We prove this relation using induction. We start by noting that the base case holds, i.e. the equation is true for $n=0$. 
    We now assume that Eq. \eqref{eq:post} holds for all values $0\le n'<n$. We will show that this implies that the equation also holds for $n$. We start from Eq. \eqref{eq:proff_start}, and write the term  with $k=0$ in the sum explicitly
    \begin{align}
        C_{2(N-n)}(\{t\}) =&\,\mel{t_1,\dots,t_{N-n}}{P_{N-n}\hat \rho_{N-n}}{t_{N-n+1},\dots,t_{2(N-n)}}
        \\\nonumber 
        &\quad + \sum_{k=1}^{n}\frac{1}{k!} \int  d\tau_1\cdots \tau_k\mel{\tau_1,\dots,\tau_k,t_1,\dots,t_{N-n}}{P_{N-n+k}\hat \rho_{N-n+k}}{t_{N-n+1},\dots,t_{2(N-n)},\tau_k,\dots,\tau_1}.
    \end{align}
    As $n-k<n$ for all terms in the sum, we can use  Eq. \eqref{eq:post} to rewrite the sum
    \begin{align}
        C_{2(N-n)}(\{t\}) =&\,P_{N-n}\mel{t_1,\dots,t_{N-n}}{\hat \rho_{N-n}}{t_{N-n+1},\dots,t_{2(N-n)}}
        \\\nonumber 
        &\quad 
        + \sum_{k=1}^{n}\frac{1}{k!} \int  d\tau_1\cdots \tau_k\sum_{\ell=0}^{n-k} \frac{(-1)^\ell}{\ell!} \int d\tau_{k+1}\cdots d\tau_{k+\ell} C_{2(N-n+k+\ell)}(\{t\}')
    \end{align}
    We now manipulate the sum of the latter terms. The sum over $\ell$ is reindexed and we change the order of summation, yielding
    \begin{align}
        C_{2(N-n)}(\{t\}) =P_{N-n}\mel{t_1,\dots,t_{N-n}}{\hat \rho_{N-n}}{t_{N-n+1},\dots,t_{2(N-n)}}
        + \sum_{\ell=1}^{n}  \int d\tau_{1}\cdots d\tau_\ell C_{2(N-n+\ell)}(\{t\}')  \sum_{k=1}^{\ell}\frac{(-1)^{\ell-k}}{k!(\ell-k)!}.
    \end{align}
    The sum over $k$ is given by $(-1)^{\ell+1}/\ell!$. Rewriting the equation, we obtain 
    \begin{align}
        P_{N-n}\mel{t_1,\dots,t_{N-n}}{\hat \rho_{N-n}}{t_{N-n+1},\dots,t_{2(N-n)}}=
        \sum_{\ell=0}^{n}\frac{(-1)^\ell}{\ell!} \int d\tau_{1}\cdots d\tau_\ell C_{2(N-n+\ell)}(\{t\}'),
    \end{align}
    which is of the same form as Eq.~\eqref{eq:post}, meaning we have proved the postulate. We reindex the expression setting $n\to N-n$ to obtain
    \begin{align}
        P_n\mel{t_1,\dots,t_n}{\hat\rho_n}{t_{n+1},\dots,t_{2n}}=
        \sum_{k=0}^{N-n}\frac{(-1)^k}{k!} \int d\tau_{1}\cdots d\tau_k C_{2(n+k)}(\{t\}').
    \end{align}
    We then take the limit $N\to\infty$, which is Eq. \eqref{eq:dmgen}.
    
    \section{Characterization parameters in terms of numerical integrals\label{sec:AppNum}}
    In this appendix, we provide the integrals required to find the characterisation parameters for the general pulsed case, where the time window $\mathcal{T}$ is set to cover the entire time range. These integrals can be used to find the characterisation parameters in sections \ref{ssec:SP} and \ref{ssec:pulse}, by inserting the form of $\chi(t)$. All parameters will be expanded to lowest order in $\chi(t)$. 
    
    The probability of success is explicitly written out to second order in $\chi(t)$,
    \begin{align}
        P_S &=  \int_{-\infty}^\infty dt_1 d\tau_1 \chi(t_1)\chi(\tau_1)e^{-\kappa |t_1-\tau_1|}+2\kappa\int_{-\infty}^\infty dt \chi(t)\int_{-\infty}^{t} dt_2 \chi(t_2)\int_{-\infty}^{t_2} dt_3 \chi(t_3)\int_{-\infty}^{t_3} dt_4e^{\kappa t_4}\int_{t_4}^{\infty} d\tau \chi(\tau)e^{-\kappa \text{max}(t,\tau)}.
    \end{align}
    Defining $\mathcal{I}_{bbn}\equiv \int_{-\infty}^\infty dt_c \ev*{b^\dagger(t_c) b(t_c)}_n$, we see that the first (second) term is $\mathcal I_{bb2}$ ($\mathcal I_{bb4}$). The density matrix components are found to second order in $\chi(t)$
    \begin{subequations}
        \begin{align}
            P_1\rho_1(t_1;t_2)
            =&\,  \frac{
                \mathcal{I}_{a0b1}(t_1,t_2)\qty(1-\mathcal{I}_{bb2}-\frac{\mathcal{I}_{bb4}}{\mathcal{I}_{bb2}})
                +\mathcal{I}_{a0b3}(t_1,t_2)+\mathcal{I}_{a2b1}(t_1,t_2)+\mathcal{I}_{a0b3}(t_2,t_1)+\mathcal{I}_{a2b1}(t_2,t_1)
            }{\mathcal I_{bb2}}\\\nonumber
            &\,+\ev*{a_1^\dagger(t_2) a_1(t_1)}
            -\frac{
                \ev*{a_1^\dagger(t_2) a_1(t_1)}\int_{-\infty}^\infty d\tau \mathcal{I}_{a1b0}(\tau,\tau)
                +\mathcal{I}_{aaabab}(t_1,t_2)+\mathcal{I}_{aaabab}(t_2,t_1)
            }{\mathcal{I}_{bb2}}\\
            P_2\rho_2(t_1,t_2;t_3,t_4)=&\,\frac{4\ev*{a_1^\dagger(t_4) a_1(t_1)}\mathcal{I}_{a0b1}(t_2,t_3)}{\mathcal I_{bb2}} 
        \end{align}
    \end{subequations}
    where we have used that $\ev*{a_1^\dagger (t)a_1(t')} = \ev*{b_1^\dagger(t) b_1(t')}$ for the considered model and we have further defined the integrals
    \begin{subequations}
        \begin{align}
            \mathcal{I}_{anbm}(t_1,t_2) &= \int_{-\infty}^\infty dt_c  \ev*{a_1^\dagger(t_2) b_0^\dagger(t_c)} \ev*{a_n(t_1)b_m(t_c)}\\
            \mathcal{I}_{aaabab}(t_1,t_2) &=  \int_{-\infty}^\infty dt_c \int_{-\infty}^\infty d\tau \ev*{a_1^\dagger(t_2) b_0^\dagger(t_c)} \ev*{a_0(\tau)b_1(t_c)}\ev*{a_1^\dagger(\tau) a_1(t_1)}.
        \end{align}
    \end{subequations}
    We now write the integrals in terms of $\chi(t)$
    \begin{subequations}
        \begin{align}\label{eq:a0b1}
            \mathcal{I}_{a0b1}(t_1,t_2) &= \kappa e^{-\frac\kappa 2 (t_1+t_2)}\int_{-\infty}^{t_1}d\tau_1\int_{-\infty}^{t_2}d\tau_2 \chi(\tau_1)\chi(\tau_2) e^{\kappa\text{min}(\tau_1,\tau_2)}=\ev*{a_1^\dagger(t_2) a_1(t_1)}\\
            \mathcal{I}_{a0b3}(t_1,t_2) &= \kappa e^{-\frac\kappa 2 (t_1+t_2)}\int_{-\infty}^{t_2}ds_1\chi(s_1)e^{\kappa s_1}\int_{-\infty}^{\infty}d\tau_1 \chi(\tau_1)\int_{-\infty}^{\tau_1} d\tau_2\chi(\tau_2)\int_{-\infty}^{\text{min}(\tau_2,t_1)}\chi(t_3)e^{\kappa \tau_3} e^{-\kappa\text{max}(s_1,\tau_1)}\\
            \mathcal{I}_{a2b1}(t_1,t_2) &= \kappa e^{-\frac\kappa 2 (t_1+t_2)}\int_{-\infty}^{t_2}ds_1\chi(s_1)e^{\kappa s_1}
            \int_{-\infty}^{\infty}d\tau_1 \chi(\tau_1)\int_{-\infty}^{t_1} d\tau_2\chi(\tau_2)\int_{-\infty}^{\tau_2}\chi(t_3)e^{\kappa \text{min}(\tau_3,\tau_1)} e^{-\kappa\text{max}(s_1,\tau_1)}\\
            \mathcal{I}_{aaabab}(t_1,t_2) &= \kappa e^{-\frac \kappa 2 (t_1+t_2)} \int_{-\infty}^{t_1}d\tau_1\int_{-\infty}^{t_2}d\tau_2 \int_{-\infty}^{\infty}d\tau_3\int_{-\infty}^{\infty}d\tau_4
            \chi(\tau_1)\chi(\tau_2)\chi(\tau_3)\chi(\tau_4)e^{\kappa[\text{min}(\tau_1,\tau_3)+ \text{min}(\tau_2,\tau_4)-\text{max}(\tau_3,\tau_4)]}.
        \end{align}
    \end{subequations}
    We then find expressions for the various parameters characterizing  the single-photon component, starting with the occupation
    \begin{align}
        P_1 = 1-\mathcal{I}_{bb2} +\frac{2\int_{-\infty}^\infty d\tau \qty[\mathcal{I}_{a0b3}(\tau,\tau)+\mathcal{I}_{a2b1}(\tau,\tau)-\mathcal{I}_{aaabab}(\tau,\tau)]-\mathcal{I}_{bb4}}{\mathcal{I}_{bb2}}
    \end{align}
    We express the purity of the one photon state, $\varpi_1,$ to zeroth order in $\chi(t)$ using the integrals  
    \begin{align}\label{eq:purGauss}
        \varpi_1&= \int_{-\infty}^{\infty} dt_1dt_2\frac{[\mathcal I_{a0b1}(t_1,t_2)]^2}{\mathcal{I}_{bb2}^2}
    \end{align}
   
    Using the relation given in Eq. \eqref{eq:a0b1}, we can write the the density matrix elements of the two-photon component   as a product of single-photon density matrix elements
     \begin{align}
        P_2\rho_2(t_1,t_2,t_3,t_4)&=4 P_S P_1^2\rho_1(t_1,t_4)\rho_1(t_2,t_3),
    \end{align}
    where $P_S$ ($\rho_{1}(t,\tau)$ and $P_1$) is taken to second (zeroth) order in $\chi(t)$.  We note that even though $\rho_2(t_1,t_2,t_3,t_4)$ is not symmetric under exchange of time arguments between $t_1$ ($t_3$) and $t_2$  ($t_4$), the density matrix element $\mel{t_1,t_2}{\rho_2}{t_3,t_4}$ is symmetric under exchange of time arguments, as can be found from by using the density matrix of the form in Eq. \eqref{eq:varrho_gen}. The relation allows us to find $P_2$
    \begin{align}
        P_2 = P_S (1+\varpi_1)
    \end{align}
    Finally, the overlap between the two-photon density matrix and combinations of eigenstates of the single-photon density matrix, $Q_{ij}$ are given to zeroth order in $x$
    \begin{subequations}
        \begin{align}
            Q_{ii} &= \frac{1}{2}\int_{-\infty}^\infty dt_1\cdots dt_4 \phi_i(t_1)\phi_i(t_2)\phi_i(t_3)\phi_i(t_4)\rho_2(t_1,t_2,t_3,t_4)=\frac{2\bra{\phi_i}\hat\rho_1\ket{\phi_i}^2}{1+\varpi_1}\\
            Q_{ij} &= \frac{1}{2}\int_{-\infty}^\infty dt_1\cdots dt_4 \phi_i(t_1)\phi_j(t_2)\phi_j(t_3)\phi_i(t_4)\left[\rho_2(t_1,t_2,t_3,t_4)+\rho_2(t_1,t_2,t_4,t_3)\right]
            =\frac{2\bra{\phi_i}\hat\rho_1\ket{\phi_i}\bra{\phi_j}\hat\rho_1\ket{\phi_j}}{1+\varpi_1}, 
        \end{align}
    \end{subequations}
    where we have used that $\ket{\phi_i}$ is an eigenstate of $\hat \rho_1$, and thus $\mel{\phi_j}{\hat \rho_1}{\phi_i}=0$.

    \section{Analytical expression for CW driving \label{sec:AppCW}}
    In this appendix, we give the analytical expressions for the CW-case. We center the time arguments around $t_c$, meaning they run from $-T/2$ to $T/2$. We will drop the subscripts referring to the detection time and the interval length for convenience.
    The density matrices components are found to second order in $x$
    \begin{subequations}
        \begin{align}\label{eq:rho1cw}
            P_{1}\rho_{1}(t_1;t_2) &= 
            \frac{\kappa}{2}  e^{-\frac{1}{2} \kappa  (| t_1| +| t_2 | )} \bigg\{1-x^2 \bigg[2+\kappa T +\left(1- e^{-\frac{\kappa  T}{2}}\right)(6+ \kappa  (| t_2 | + | t_1| +T)
   \bigg]\bigg\}\\\nonumber&
   + \frac{\kappa x^2}{2} e^{-\frac{1}{2} \kappa  (-| t_1| +| t_2 |+T)} \left[3+\kappa\left(\frac{T}{2}-|t_1|\right)\right]
   + \frac{\kappa x^2}{2} e^{-\frac{1}{2} \kappa  (| t_1| -| t_2 |+T)} \left[3+\kappa\left(\frac{T}{2}-|t_2|\right)\right]\\\nonumber&+\kappa x^2 e^{-\frac{1}{2}
   \kappa  (| t_1-t_2 | +T)}(2+\kappa  | t_1-t_2 |) 
   \\    
            P_{2}\rho_{2}(t_1,t_2;t_3,t_4)&=4x^2\kappa^2 e^{-\frac \kappa 2(\abs{t_1-t_3}+\abs{t_2}+\abs{t_4})}\qty(1+\frac{\kappa\abs{t_1-t_3}}{2}).
        \end{align}
    \end{subequations}
    From the density matrices, we obtain the population of the different density matrix components
    \begin{subequations}\label{eq:probscw}
        \begin{align}
            P_{1}&= 
            1-e^{-\frac{\kappa T}{2}} -x^2\bqty{10+2\kappa T-e^{-\frac{\kappa T}{2}}\qty(18+9\kappa T+\frac{\kappa^2 T^2}{4})+2e^{-\kappa T}\qty(4+\kappa T)}\\
            P_{2} &= x^2\qty{
                10+2\kappa T-e^{-\frac{\kappa  T}{2}} \left(\frac{\kappa ^2 T^2}{4} +6 \kappa  T+14\right)+e^{-\kappa T} (\kappa  T+4)
            }.
        \end{align}
    \end{subequations}
    Furthermore, the purity of the one photon state, which is equivalent to the HOM visibility when $P_2=0$, is given by
    \begin{align}
        \varpi_1 = 1 - 4x^2\frac{e^{-\frac{\kappa T}{2}} \bqty{
        \kappa T-5+e^{-\frac{\kappa T}{2}}\pqty{7+\kappa T + \frac{(\kappa T)^2}{8}}-e^{-\kappa T}\pqty{2+\frac{\kappa T}{2}}}}{\pqty{1 - e^{-\frac{\kappa T}{2}}}^2},
    \end{align}
    and the population of the two-photon density matrix with two photons in the state $\ket{\phi_\text{CW}}$ is
    \begin{align}
        Q_{11}= \frac{
            2 \qty{40-e^{-\frac{\kappa T}{2}} \bqty{(\kappa T)^2+16 \kappa T+56}+4e^{-\kappa T}\pqty{ \kappa T+4}}
        }{
            8 (\kappa T+5)-e^{-\frac{\kappa T}{2}} \bqty{(\kappa T)^2+24 \kappa T+56}+4e^{-\kappa T} (\kappa T+4)
            }.
    \end{align}
\end{widetext}

\end{document}